\documentclass[epj]{svjour}
\usepackage{graphics}
\usepackage{psfig}
\begin{document}
\title{Jahn-Teller, Charge and Magnetic Ordering in half-doped
  Manganese Oxides}
\author{S. Fratini\inst{1}, D. Feinberg\inst{2} and M. Grilli\inst{1}}

%
\offprints{}          
\institute{$^1$ Istituto Nazionale di Fisica della Materia and
  Dipartimento di Fisica, Universit\`a di Roma "La Sapienza", Piazzale
  Aldo
  Moro 2, 00185 Roma, Italy\\
  $^2$ Laboratoire d'Etudes des Propri\'et\'es Electroniques des
  Solides, Centre National de la Recherche Scientifique, associated
  with Universit\'e Joseph Fourier, BP 166, 38042 Grenoble Cedex 9,
  France}
\date{Received: date / Revised version: date}
\abstract{ The phase diagram of half-doped manganite systems of
  formula A$_{0.5}$A$^\prime_{0.5}$MnO$_3$ is investigated within a
  single-orbital model incorporating magnetic double-exchange and
  superexchange, together with intersite Coulomb and electron-lattice
  interactions.  Strong Jahn-Teller and breathing mode deformations
  compete together and result in shear lattice deformations.  The
  latter stabilize the charge-ordered CE-type phase, which undergo
  first-order transitions with temperature or magnetic field to either
  Ferromagnetic metallic or Paramagnetic insulating phases.  An
  essential feature is the self-consistent screening of Coulomb and
  electron-phonon interactions in the ferromagnetic phase.
} 
\authorrunning{S. Fratini {\it et al.},} \titlerunning{Jahn-Teller, charge 
and magnetic ordering in half-doped manganese oxides} \maketitle
%

Manganese perovskite oxides are currently the object of intense
activity.  Motivated initially by the colossal magnetoresistance
phenomena, more recent studies have revealed an extremely rich phase
diagram originating from the interplay of charge, lattice, orbital and
magnetic degrees of freedom \cite{rev}.  The general formula is
A$_{1-x}$A$^\prime_{x}$MnO$_3$ where A is in general a trivalent rare
earth element (La, Pr, Nd) and A$^\prime$ a divalent alcaline element
(Sr, Ca). Substitutional doping allows to explore the full phase
diagram, from $x = 0$ to $x = 1$.  At the extremes, LaMnO$_3$ and
CaMnO$_3$ are antiferromagnetic insulators.  The former is a layered
antiferromagnet, which can be explained thanks to the large
Jahn-Teller couplings of the $e_g$ electrons of Mn$^{3+}$ ions
\cite{us}. The latter shows a N\'eel ordering due to antiferromagnetic
exchange of $t_{2g}$ electrons \cite{wollan}.  With doping, the
double-exchange phenomena originating from Hund's coupling between
$e_g$ and $t_{2g}$ electron spins can stabilize a metallic
ferromagnetic phase \cite{Zen,And,DeGen}: Coherent band motion occurs
for ferromagnetic ordering, while strong inelastic scattering takes
place in the high temperature paramagnetic phase.  Very large
magnetoresistance is obtained when the applied magnetic field is able
to align $t_{2g}$ spins, thereby favouring the metallic phase.
Nevertheless, it has been pointed out that spin scattering alone is
not sufficient to quantitatively explain the phenomenon.  Millis {\it et al.}
\cite{Millis} suggested that a large electron-lattice coupling is
involved, with the formation of Jahn-Teller polarons in the insulating
phase.  Such large couplings are quite expected from the very large
cooperative Jahn-Teller distortions existing in LaMnO$_3$. Those
deformations indeed involve more than ten per cent variations of the
$Mn-O$ bond lengths around all Mn$^{3+}$ ions. Local deformations have
been indeed revealed in charge-disordered phases by X-ray and neutron
spectroscopy, as well as optical measurements.  They consist of
Jahn-Teller deformations around Mn$^{3+}$ ions, and "breathing mode"
deformations with shorter $Mn-O$ bonds around Mn$^{4+}$ ions.  The
role of these deformations becomes more stringent in the
charge-ordered phases (CO) of doped manganites. These phases strongly
compete with the ferromagnetic metallic (FM) one at sufficient doping.
Besides the Coulomb interaction between electrons on Mn ions,
electron-phonon interaction should play a prominent role in this
phenomenon.  This is exemplified by the nature of charge ordering at
half-doping, for instance in La$_{0.5}$Ca$_{0.5}$MnO$_3$: While Mn$^{3+}$
and Mn$^{4+}$ ions alternate in two directions (say, a and b), in the
other direction (which we here define as the c-axis), one finds rows
of Mn$^{3+}$ or Mn$^{4+}$ ions.  If CO were exclusively due to
intersite Coulomb interaction, one would on the contrary expect a
Wigner crystal ordering, alternated in all directions.  This shows
that cooperative lattice distortions are an essential ingredient to
understand charge ordering \cite{yunoki}.

Charge ordering at $x = 0.5$ is accompanied by CE-type
antiferromagnetic order: In the ab directions, it involves
ferromagnetic and antiferromagnetic zigzag chains crossing each other.
A qualitative explanation was given a long time ago by Goodenough
\cite{Good}, following the pioneering structural analysis of Wollan
and Koehler \cite{wollan}: The cooperative Jahn-Teller distortions
are accompanied by orbital ordering, and induce the magnetic
structure. Moreover, away from half-doping, this CE structure appears
as an elementary "brick" to build more complicated charge ordering
patterns such as "stripes" \cite{mori}.  It is thus especially robust
and calls for a detailed explanation.

A few models have been proposed to explain CE ordering, putting the
emphasis either on intersite Coulomb interactions \cite{pandit},
magnetism and orbital ordering \cite{solovyev,jackeli}. Mizokawa et
al.  \cite{mizokawa}, and Yunoki and coworkers \cite{yunoki} have
underlined the prominent role of Jahn-Teller deformations.

Let us first list and grossly estimate the various energy scales in
the system.  The on-site Hubbard repulsion $U$ and the 
atomic level difference between the $e_g$ orbitals of manganese 
and the $2p$ orbitals of the oxygen are of the order of several
$eV$'s, and are larger than the total conduction bandwidth ($W
\sim 3 eV$).  The Hund coupling $J_H$ is of order $1 eV$, while the
intersite Coulomb repulsion seems not to be larger than $0.5 eV$.  The
Jahn-Teller splitting in the insulating LaMnO$_3$ phase is comparable,
as shown by spectroscopy and optical absorption measurements
\cite{dessaushen,Jung}. In terms of a local electron-phonon coupling,
it is reasonable to think of energies of the order of $0.2-0.3 eV$,
comparable to the intersite $e_g$ hopping integrals $t_0 \sim 0.1-0.4
eV$ depending on the d-orbitals involved.  On the other hand, the
magnetic couplings (which in a cubic lattice give rise to critical
temperatures $T_c$ between $100K$ and $400K$) are in the range of a
few $meV$.  This holds as well for the superexchange
(antiferromagnetic) couplings as, more surprisingly, for the
(ferromagnetic) double-exchange ones.  It has been shown by Zener
\cite{Zen} that $T_c^{DE} \sim \alpha t_0$, {\it e.g.} is proportional to
the total kinetic energy of the carriers.  As will be shown below,
$\alpha$ is quite small and the actual values of $T_c^{DE}$ can be
easily explained with a realistic $t_0$, for instance within De
Gennes's mean-field picture \cite{DeGen}.  This hierarchy of energy
scales is completed by the one set by the external magnetic field
needed to turn the FM phase into the CE (AFCO) phase: It ranges from
a few Teslas to $20$ Teslas or more.  In terms of energy scale per
atom, this is very small, of the order of $0.4-4 meV$.  It is thus
consistent with the values of the magnetic exchange constants, but
much smaller than all the other scales. This points towards an
important conclusion: The stringent competition between the above
phases require that their free energies be very close, in the range of
a few $meV$ per atom.  Owing to the much larger electron-phonon and
Coulomb interactions, it is reasonable to suppose that they play a
dominant role in stabilizing the low-temperature CE phase. The
necessary conclusion is that CE and FM phases are (meta)stable minima
of the free energy, separated by rather high barriers. This is
consistent with the fact that the phase transitions (with temperature
or magnetic field) between charge ordered and charge disordered phases
are first-order, with strong hysteresis under magnetic field.
Tendencies to phase separation between FM and CO phases have been
demonstrated in La$_{0.5}$Ca$_{0.5}$MnO$_3$,
Pr$_{0.7}$Sr$_{0.3}$MnO$_3$ and other compositions.  One should also
notice that charge ordering is always strong when it exists.  Fine
tuning of the chemical composition between CO and FM low temperature
phases \cite{tokura2} does not allow to stabilize "weak" charge
ordering.  This points towards strong interactions
(electron-phonon or Coulomb) in the insulating phase, while they are
screened in the metallic phase.  This feature is overlooked by
mean-field treatments, but can be recovered by taking into account
exchange-correla\-tion corrections to the intersite Coulomb repulsion,
as shown by Sheng and Ting \cite{sheng}. Since the lattice distortions
here also come from Coulomb interactions (between $Mn$ and $O$ ions),
we propose here to generalize the screening idea to electron-phonon
interactions and use for this purpose a phenomenological approach.

Given the complexity of the overall Hamiltonian, here we restrict
ourselves to a single-orbital model in two dimensions, which
quantitatively reproduces the various phase diagrams and their tuning
by subtle variations of the bandwidth.  Our goals are i) obtaining,
for realistic values of the parameters, FM, CE and paramagnetic
phases; ii) exploring by small variations of those parameters the
different kinds of phase diagrams, with temperature and magnetic
field: Of the type of La$_{0.5}$Sr$_{0.5}$MnO$_3$ (no charge
ordering, FM-PM transition with increasing $T$); of the type of
Nd$_{0.5}$Sr$_{0.5}$MnO$_3$ (CE-FM-PM transitions with $T$, CE-FM with
$H$); of the type of Pr$_{0.5}$Ca$_{0.5}$MnO$_3$ (CE-PMCO-PM
transitions with $T$, CE-FM with $H$).  iii) obtaining first-order
transitions between CE and FM phases.

Taking into account explicitly orbital ordering should not change
qualitatively the results since it works in the same direction
\cite{yunoki,jackeli} but may lead to quantitative improvement.

\section{Model and approximations}

\subsection{Hamiltonian}

According to the arguments given in the introduction, we assume an
infinite repulsion ($U=\infty$) between electrons on the same lattice
site, and an infinite Hund coupling ($J_H=\infty$) between the
localized $t_{2g}$ spins and the itinerant $e_g$ spins. One can
therefore consider spinless electrons, their spin degree of freedom
being unequivocally defined by the direction of the local $t_{2g}$
spins $\vec{S}$.  Furthermore, we consider in this work a
two-dimensional plane of the structure, with a half-filled band made
of a single $e_{g}$ orbital.  The effective model Hamiltonian is then:
\begin{equation} \label{H}
H= H_{DE}+H_{Coul}+H_{ph}+H_{SE}+H_{H}
\end{equation}
with
\begin{eqnarray*}
H_{DE}&=& - \sum_{<ij>} \tilde{t}_{ij} c^\dagger_{i} c_{j} \\
H_{Coul} &=& \sum_{<ij>} V (n_i-n) (n_j-n)\\
H_{ph}&=&  \frac{1}{2} \sum_i [K_b Q_{bi}^2 +K_2 Q_{2i}^2 +K_s
Q_{si}^2 ] \\ && \hspace{-1.5cm}
 -\sum_i g_2 Q_{2i} (n_i-n) + \sum_i g_b Q_{bi} (n_i-n)
- L_s \sum_{<ij>} Q_{si} Q_{2j}  \\
H_{SE}&=& \sum_{<ij>} [J_1-J_2 Q_s] \vec{S}_i\cdot \vec{S}_j \\
H_{H}&=& -g\mu_{b}\vec{H} \sum_{i}\vec{S}_i
\end{eqnarray*}

The first term $H_{DE}$ represents the double exchange hopping of
electrons on a square lattice.  Here $c^\dagger_{i}$, $c_{i}$ are
respectively the creation and annihilation operators for spinless
electrons from a single band, and $\tilde{t}_{ij}=t \cos
(\theta_{ij}/2)$ is the transfer integral between neighboring Mn sites
whose ionic spins $\mathbf{S}_i$ and $\mathbf{S}_j$ make an angle
$\theta_{ij}$ \cite{DeGen}.

The second term $H_{Coul}$ describes the Coulomb repulsion between
nearest neighbors ($n_i=c^\dagger_i c_i$ and $n$ is the average
electron density, which is equal to $1/2$ in the present case).

The third term $H_{ph}$ is the elastic part, which includes the
coupling of electrons to a Jahn-Teller (JT) mode $Q_2$ and of holes to
a ``breathing'' mode $Q_b$ ($g_2$ and $g_b$ are the coupling
strengths, $K_2$ and $K_b$ the spring constants).  In the planar
geometry considered here, the other Jahn-Teller mode $Q_1$ is not
relevant.  We have also introduced a shear mode $Q_s$, which is driven
by $Q_2$. Such a shear deformation, which is experimentally observed
at low temperatures, is essential to reconcile the alternating
Mn$^{4+}$ breathing and Mn$^{3+}$ JT distortions wich develop in the
ordered phases. A substantial shear deformation is indeed observed in
La$_{0.5}$Ca$_{0.5}$MnO$_3$ \cite{radaelli}.  It results in some
$Mn-O-Mn$ bonds being shorter and other larger ("zig-zag" chains, see
Fig. 1).

The term $H_{SE}$ represents the antiferromagnetic (AF) superexchange
interaction $J_1$ between the
ionic spins
on neighboring sites, which are treated as classical. The additional
term $J_2Q_s$ is a phenomenological implementation of the Goodenough
rule: It can either enhance or reduce the AF coupling depending on
the sign of the shear deformation, which accounts for the fact that
longer (shorter) Mn-Mn bonds have a more (less) antiferromagnetic
character \cite{Good}.  The last term $H_{H}$ takes into account the
external magnetic field.

We shall study the Hamiltonian (\ref{H}) in the mean-field
approximation, describing the charge ordered (CO) phase as a charge
density wave (CDW) with momentum $(\pi,\pi)$.  Let us call $\bar{n}^A$
and $\bar{n}^B$ the average electron densities in the two resulting
sublattices, which correspond respectively to the Mn$^{3+}$ and
Mn$^{4+}$ ions.  We shall further assume that the JT coupling is only
active on A sites, while the breathing deformations arise on B sites.
With these approximations, the terms in the Hamiltonian which depend
explicitely on $(n_i-n)$ reduce to
\begin{equation}
H_{MF}=-\Delta \sum_{i \in A or B} (n_i^A-n_i^B) + const
\end{equation}
where the order parameter $\Delta$ is defined as
\begin{equation}
\label{Delta}
\Delta=2V(\bar{n}^A-\bar{n}^B) + (g_bQ_b+g_2Q_2)/2
\end{equation}
and the chemical potential has been set to zero by adding a term $
\Delta\mu=-(g_2Q_2-g_bQ_b)/2 $ to recover particle-hole symmetry (with
these notations, the choice $\bar{n}^A \ge \bar{n}^B$ corresponds to
the $Q$'s being all positive).

The magnetic part is also treated in mean-field, according to de
Gennes' procedure \cite{DeGen}, using a gaussian distribution for the
angle of the classical spins with respect to the mean field direction.
We consider the following magnetic phases: Ferromagnetic (F),
paramagnetic (P), N\'eel anti-ferromagnetic (NAF), and CE-type
ordering, with ferromagnetic zig-zag chains, coupled
anti-ferro\-magne\-ti\-cally (CE).  The most general unit cell which allows
to describe all these phases is is made of 8 nonequivalent Mn sites in
a plane (Fig. 1). In each of these magnetic configurations, the total
free energy is minimized with respect to the following parameters: i)
the magnetization on non-equivalent magnetic sites, ii) the average
electron density $\bar{n}^A$ on sublattice A ($\bar{n}^B$ being just
$1-\bar{n}^A$) and iii) the lattice displacements.

\begin{figure}
  \resizebox{9cm}{!}{\includegraphics{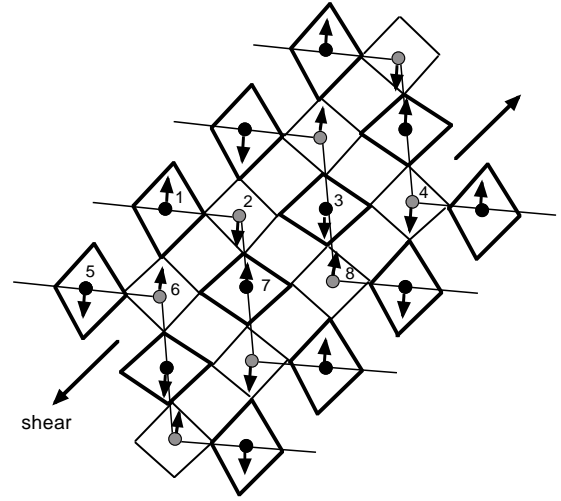}}
\caption{
  The CE structure and the unit cell (ions 1 to 8) in the ab plane: $Mn^{3+}$
  Jahn-Teller ions (black circles) are at the center of distorted
  $Mn-O$ rhombus, and $Mn^{4+}$ ions (grey circles) undergo shear
  displacements, forming zig-zag chains with short (antiferromagnetic)
  and long (ferromagnetic) bonds.}
\label{Fig1}
\end{figure}

\subsection{Phenomenological treatment of screening effects}

As mentioned in the Introduction, to have a realistic description of
the phase diagram which includes both the metallic and charge ordered
phases at half-filling, it is necessary to go beyond the simple
mean-field approach described in the preceding section by including
the effects of exchange and correlation. Such effects on the intersite
Coulomb repulsion in the half-doping manganites have recently been
analyzed within an RPA-like calculation \cite{sheng}, which is known
to be appropriate for interacting electron systems at metallic
densities.  Since a detailed study goes beyond the scope of this
paper, we propose here a semi-phenomenological treatment of screening
which allows a qualitative description of the transition between CO
and metallic states, and which correctly reproduces the results of
reference \cite{sheng}. The method is further generalized to describe
the screening of the electron-lattice interactions. Actually, the
latter are due to the Coulomb repulsion between $Mn$ and $O$ ions and
are therefore also screened in the metallic phase. This screening
should be weaker than that of $Mn-Mn$ interactions, since it involves
$Mn-O$ rather than $Mn-Mn$ charge fluctuations, but it should be
sizeable.

The procedure will be carried out in two successive steps.  The first
step consists in writing a reasonable estimate for the
exchange-correlation energy $E_{xc}$, which is defined as the
correction to the ground-state energy beyond the Hartree mean-field
result. In the second step, we shall define an effective hamiltonian
$H_{xc}$ to be treated in the mean-field approximation, such that
\begin{equation}
\label{Hav}
\langle H_{xc} \rangle = E_{xc}
\end{equation}
This results in a modification of the atomic energy levels $\pm
\Delta$ (i.e. in a reduction of the CDW gap), and it yields a
correction to the free energy which is precisely of the form $E_{xc}$.

\subsubsection{Exchange-correlation energy}

Let us start by analyzing the simple ferromagnetic case at $T=0$,
where the electron hopping is not renormalized by the mechanism of
double exchange.  In the metallic phase, which corresponds to a
vanishing order parameter $\Delta$, the leading correction comes from
the exchange (Fock) terms. These terms are responsible for an increase
of the carrier itinerancy, which can be viewed as a renormalization of
the hopping parameter $t\rightarrow t+V\langle
c^\dagger_ic_{i+\delta}\rangle $. Hence the kinetic energy is lowered
by a quantity proportional to the interaction potential $V$, and one
can write $E_{xc}= -a V$ (the parameter $a$ is related to the
dielectric constant of the system).  On the other hand, in the CO
phase, i.e.  at strong $\Delta$, the correlation energy corresponds to
the interaction between density fluctuations on neighboring sites,
each of them being proportional to $\delta n \sim t/\Delta$.
Therefore, in this case the appropriate limiting formula is
$E_{xc}\sim -V (t/\Delta)^2$.  These results can be generalized to the
screening of the electron-lattice interactions, by replacing
$V\rightarrow g^2/K$ and by introducing the corresponding order
parameter $\Delta$ as given by eq. (\ref{Delta}).

A smooth interpolation between the weak and strong coupling behaviors
can be obtained by writing the following expression for the
exchange-correlation energy:
\begin{equation} \label{Exc}
E_{xc} = - \frac{aV+b(g_b^2/K_b+g_2^2/K_2)}
{1+c \left( \frac{\Delta}{t}\right)^2}
\end{equation}
where $a,b$ and $c$ are phenomenological parameters (the ratio
$a/c=1.44$ can be deduced from ref. \cite{sheng} and $b/a=1/10$ is
chosen according to the ionic distances).

As was stated at the beginning of this section, this formula is only
appropriate in the ferromagnetic case. It does not account for the
fact that the mobility of the carriers taking part in the screening
process is affected by the magnetic structure through the DE
mechanism.  We shall give here the arguments which allow a
generalization of eq. (\ref{Exc}) to the different kinds of magnetic
orderings.

In the free-electron limit ($\Delta\rightarrow 0$), where screening is
due to coherent band motion, one expects the correlation energy to be
reduced by a factor $\tilde{t}/t$, where
\begin{equation}
\label{teff}
\tilde{t}=t \langle \cos(\theta_{ij}/2) \rangle
\end{equation}
is the effective DE hopping parameter averaged in all space directions
(this gives respectively $1,8/9,1/2$ in the $F,P$, $CE$ phases).  The
situation is slightly more complicated in the charge ordered phases,
because $E_{xc}$ comes from incoherent hopping of the carriers to
neighboring sites.  According to Hund's rule, such processes will be
allowed only between sites with parallel spins, which defines an
effective number of neighbors $\tilde{z}\le z$.

In the CE phase, for instance, the lattice can be divided into U (up)
and D (down) sites, according to the spin direction.  
Since each site has 2 U and 2 D neighbors, a given
electron can only hop to the 2 neighbors with the same spin direction,
and consequently $\tilde{z}/z=1/2$.
At finite temperatures, however, thermal fluctuations will reduce the
absolute value of the local magnetization $m$. Accordingly, there will
be a finite probability that a given U site has a $\downarrow$ spin,
which is given by $n_U^{\downarrow}=(1-m_U)/2$ (an equivalent
expression holds for D sites).  The total probability for hopping away
from a U site is therefore
\[
2 n_U^\uparrow n_U^\uparrow +2 n_U^\downarrow n_U^\downarrow +
2 n_U^\uparrow n_D^\uparrow +2 n_U^\downarrow n_D^\downarrow
\]
where obviously $n_U^{\uparrow}=1-n_U^{\downarrow}$.  By adding the
contributions for hopping processes starting from both U and D sites
and dividing by 2, we obtain
\begin{equation}
\label{zeff}
\tilde{z}=\frac{1}{2} \left[ 4+ (m_U+m_D)^2\right]
\end{equation}
which correctly gives $\tilde{z}/z=1,1/2,1/2$ for the F,P,CE phases at
$T=0$.  Here the factors (\ref{teff}) and (\ref{zeff}) introduce a
feedback on the itinerancy of the electrons in the case of an applied
magnetic field, which tends to align the spins ferromagnetically. This
effect is essential in reducing the critical $H$ for the CE-FM
transition at low temperatures, as we shall see below.

For each given magnetic configuration, instead of eq. (\ref{Exc}), we
shall use the following formula for the exchange-correlation energy:
\begin{equation}
\tilde{E}_{xc} =
- \frac{\left[\tilde{a}V+\tilde{b}(g_b^2/K_b+g_2^2/K_2)\right]}
{1+\tilde{c} \left( \frac{\Delta}{t}\right)^2}
\end{equation}
where the screening parameters $a,b,c$ have been modified according to
\begin{equation}
\tilde a= a \frac{\tilde{t}}{t}, \hspace{1cm} \tilde{b}= b
\frac{\tilde{t}}{t}, \hspace{1cm}
\tilde{c}= c \frac{z}{\tilde{z}}\frac{\tilde{t}}{t}
\end{equation}
We emphasize here that the terms in the numerator of Eq. (\ref{Exc})
are rescaled by the $\tilde{t} /t$ factor since they arise from the
coherent screening processes (mostly active when $\Delta \to 0$).
On the other hand the local (incoherent) screening processes
related to the term in the denominator of Eq. (\ref{Exc})
are also rescaled by the effective number of accessible nearest neighbor sites.

\subsubsection{Mean-field potential from exchange and correlation}

We wish to define an effective hamiltonian $H_{xc}$ to be treated in
the mean-field approximation such that the correction to the free
energy is equal to $E_{xc}$.
To this purpose, we replace $\Delta$ by an operator $\hat{\Delta}$
(e.g. the mean-field parameter $n^A$ is replaced by $\sum_{i \in A}
n^A_i$), and linearize the resulting expression.  This gives
\begin{eqnarray}
\label{Hexop}
H_{xc}=  B
\tilde{c} \frac{V}{t}\frac{\Delta}{t} \sum_{i\in A or B} (n^A_i-n^B_i)
+ const
\end{eqnarray}
where we have defined
\begin{equation}
B=
\left[\tilde{a}V+\tilde{b}(g_b^2/K_b+g_2^2/K_2)\right]/
\left[1+\tilde{c} \left( \frac{\Delta}{t}\right)^2\right]^2
\end{equation}
The constant part in eq. (\ref{Hexop}) is
\begin{equation}
-B \left\lbrace 1+ \tilde{c}  \frac{\Delta}{t^2}
\left[ 6V(\bar{n}^A-\bar{n}^B) + (g_bQ_b+g_2Q_2)/2 \right]
\right\rbrace
\end{equation}
It is easy to verify that eq. (\ref{Hav}) holds when $\bar{n}^A =
\langle n^A \rangle$ and $\bar{n}^B = \langle n^B \rangle$.

One notices that a dielectric constant can be deduced from the
screening of the gap, by writing
\[
\Delta \rightarrow \Delta_{eff}=\Delta -B\tilde{c}\frac{V\Delta}{t^2}
\]
which gives
\[
\varepsilon=\frac{\Delta}{\Delta_{eff}}=\frac{1}{1-c \frac{V}{t^2}
  \frac{\tilde{a}V+\tilde{b}(g_b^2/K_b+g_2^2/K_2)}{ 1+\tilde{c} \left(
      \frac{\Delta}{t}\right)^2}}
\]

\section{Results}

\subsection{The phase diagram: Existence of a CE phase}
The Hamiltonian in Eq. (\ref{H}) is formed by several competing terms,
and the corresponding phase diagram contains several phases, each one
dominating in some region of the parameter space. To make the analysis
simpler, we choose to vary together those parameters having similar
physical effects. In particular, the electron-phonon couplings
generically reduce the electron mobility and, at mean-field level,
tend to give rise to a staggered charge ordering, acting similarly to
the n.n. electron-electron repulsion $V$.  Therefore in varying $V$ we
keep constant the ratio $V/(g^2/K)$. For the sake of simplicity, we
also keep a fixed $J_2/J_1$ ratio, although it is not the only
possible choice.
\begin{figure*}
  \centerline{\resizebox{16cm}{!}{\includegraphics{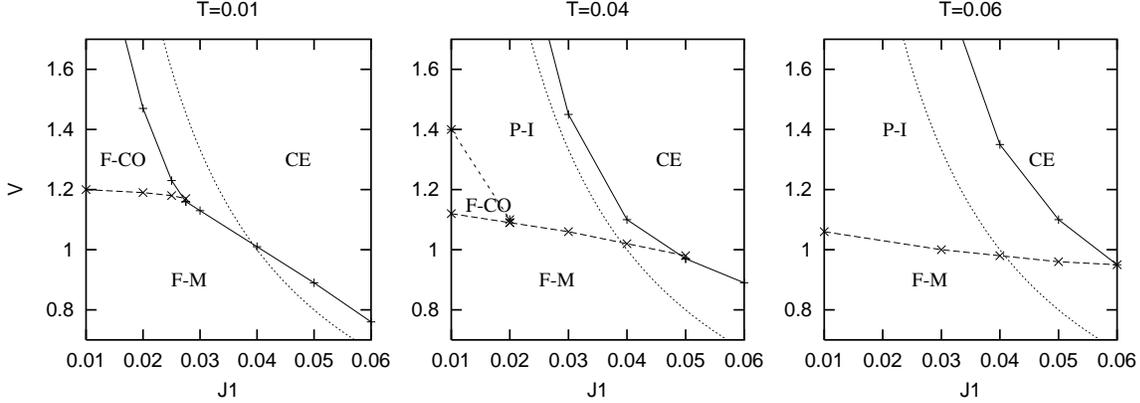}}}
\caption{
  Phase diagrams as function of the AF coupling $J_{1}$ and the
  electron-phonon or Coulomb parameter $V$, for different temperatures
  (see text).}
\label{Fig2}
\end{figure*}

Figure 2 reports typical phase diagrams at various temperatures
as a function of the magnetic coupling $J = J_1 S^{2}/t$ with $J_{1} =
J_{2}$ ($S = 3/2$) and of the repulsive e-e interaction
$V=0.5(g^2/K)$. At low temperature (left panel) there is a metallic (i.e.
without charge-ordering) ferromagnetic phase (FM) when the
charge-ordering (CO) terms $V$ and $g^2/K$ are not too large. This FM
phase is naturally suppressed by the increase of the antiferromagnetic
(AF) superexchange coupling $J_1$. When the charge mobility is
suppressed by the CO terms, one finds two distinct possible phases. At
low values of the AF coupling the pure CO effects dominate and a
ferromagnetic (F) CO phase occurs at sufficiently large values of $V$
(F-CO).  The transition is first order, as found in Ref \cite{sheng},
due to the exchange-correlation terms.  On the other hand, by
increasing the AF coupling, the CO ferromagnetism is destabilized and
a CE phase takes place.  This latter phase naturally realizes the best
compromise between the electron mobility, favored by the ferromagnetic
bonds, the CO, and the AF interactions increasing with $J_1$.  The CE
ordering arises due to competing lattice displacements.  In particular
a substantial shear mode is induced in the lattice to reconcile the
(asymmetrical) JT deformations occurring in the Mn$^{3+}$ ions with
the (centrosymmetric) breathing deformations around the Mn$^{4+}$
ions.  The resulting lattice structure displays zig-zag chains formed
by long bonds interlaced with zig-zag chains of short bonds.  Then the
peculiar CE magnetic structure naturally appears.  In particular,
according to Goodenough \cite{Good}, orbital ordering makes the sign of
the magnetic couplings to be correlated to the length of the bonds,
with AF (F) magnetic couplings corresponding to short (long) bonds.
Therefore, the lattice-driven chains with short bonds and with long
bonds naturally translate into a lattice-driven CE magnetic structure.

The temperature evolution is represented in the three panels of figure 2.
By increasing $T$ the weak ferromagnetism surviving in the CO phase at
small $J$ is rapidly suppressed in favor of a a CO paramagnetic phase.
Also the CE region, being due to a delicate balance between CO, FM,
and AF interactions, is reduced rather rapidly. The FM phase at small
values of $J_1$ is instead based on the double-exchange mechanism,
which is more robust and, upon increasing $T$, is only weakly
``invaded'' by the CO paramagnetic phase.  One also observes that
compounds with a low-temperature CE magnetic structure, but laying
near the CE-FM phase boundary, can undergo a first-order CE-to-FM
transition upon increasing the temperature.

It is worth pointing out that, within our formal scheme, the CO
paramagnetic phase is the only possible mean-field description of an
insulating non-magnetic phase at moderate temperature values.  The
explored temperature range is indeed much too low to allow for a
thermal disruption of the CO, which instead occurs at temperatures of
the order of $V$. However, in strong coupling, one may speculate that
a more refined description would allow for the disordering of the
charge (possibly without spoiling the local lattice deformations
around the charges) thus producing the disordered paramagnetic (and
polaronic) phase which is observed in all manganites above a few
hundreds of kelvins.

\subsection{Sensitive role of the electronic bandwidth}
A key issue is the role of the kinetic energy in the competition
between the different phases. An extended experimental analysis of the
phase diagrams of the various manganites \cite{tokura,tokura2}
suggests that the electronic bandwidth, among other parameters such as
lattice disorder, plays a primary role in determining the stability
and the competition between the FM and the insulating phases. In our
model we investigated this relevant point by varying the bare hopping
amplitude in front of the double-exchange term in the Hamiltonian
(\ref{H}).  We also assume that the same mechanism inducing the
variation of the nearest neighbour hopping of the itinerant electrons
in the $e_g$ orbitals is responsible for variations of the hopping of
the $t_{2g}$ electrons as well. This affects the superexchange
couplings $J_1$ and $J_2$, in particular $J_1$ is expected to arise
from second-order hopping processes of the $t_{2g}$ electrons
$J_1\propto t^2/U_{t_{2g}}$ (where $U_{t_{2g}}$ is some effective
repulsion between electrons in the same doubly occupied $t_{2g}$
orbital).  According to this assumption, when the hopping $t$ is
increased without changing the intersite repulsion $V$ one moves
downwards in the phase diagrams of Fig. 2, where the variable $V/t$ is
reported on the $y$-axis. At the same time, however, the increasing
$t$ produces an increase in the $x$-axis variable $J \propto t$.
Therefore, by keeping all the Hamiltonian parameters fixed, but $t$
and $J_1=J_2$, one moves in the phase diagram along the dotted curves
$V/t=A/J$ reported in Fig. 2. These curves correspond to similar
physical systems, where the only nearest neighbour electronic hopping
amplitudes have been varied.
\begin{figure*}
  \centerline{\resizebox{13cm}{!}{\includegraphics{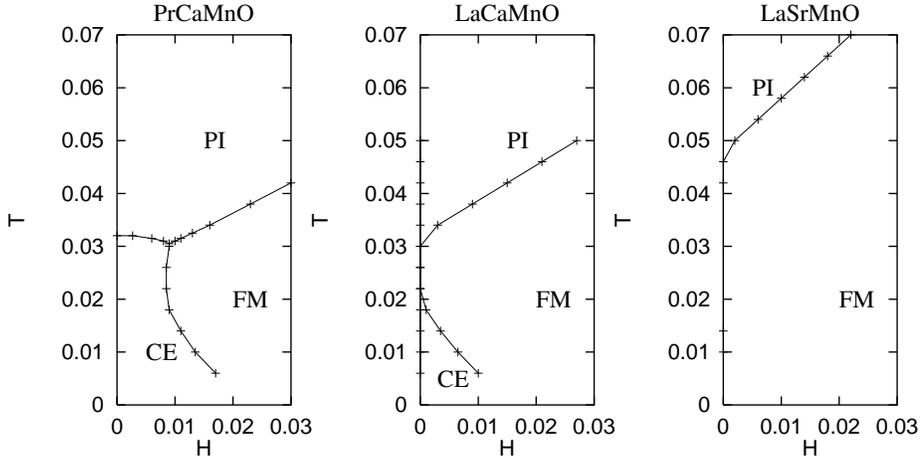}}}
\caption{Phase diagrams as function of temperature and magnetic
  field, for three typical bandwidth parameters (see text).}
\label{Fig. 3}
\end{figure*}

It immediately appears that systems which slightly differ by the
electronic hopping amplitude can have different magnetic structures at
low temperature (Fig. 2, left panel). In particular, by increasing $t$ a
first-order transition can occur at low temperature from a CE to a FM
phase.  Furthermore, the temperature evolution of systems with
different bandwidth but laying near a phase boundary can be different.
This is made apparent in Fig. 3, where we report the
temperature-magnetic field phase diagrams for three systems with
different hopping (and therefore also different magnetic couplings
$J_1$) laying on the same dotted curve $(V/t)=0.04/J$. The three phase
diagrams correspond to systems with slightly different magnetic
couplings (differing at most by ten percent).  Nevertheless, already
at zero magnetic field, the three systems display a completely
different evolution in temperature.  The more insulating (i.e., with
smaller $t$) system having $J=0.037$ (so that $(V/t)=1.081$) never
becomes metallic at zero field, but only undergoes a first-order
transition from a low-temperature CE phase to a paramagnetic
insulating phase at $T\simeq 0.032 t$ (left panel, see also Figs.4,5).
In the more metallic
system (center panel)  having $J=0.038$ (and $V/t=1.053$), the CE
phase disappears at a lower temperature $T\simeq 0.02 t$ and it is
replaced by an intermediate FM phase.  The ferromagnetic order and the
metallicity is then destroyed at a higher temperature $T\simeq 0.032t$
where a paramagnetic CO phase takes place. We reiterate here that this
latter phase is better to be intended as the mean-field representation
of a disordered paramagnetic insulating phase. Finally, at even larger
values of $J=0.04$, corresponding to $V/t=1$ the metallic phase is
present already at low temperature and it survives up to a $T\simeq
0.05 t$.

The relevant role of the kinetic energy in stabilizing the uniform
metallic FM phase at the expenses of the CO phases and particularly of
the one with CE magnetic order is made even more apparent in the
presence of a magnetic field. This is particularly clear in the first panel
of Fig. 3, where the FM phase, which would be absent at zero field
becomes the most stable solution at large enough $H$.

It is also worth mentioning that, due to the presence of screening,
the metallic uniform solution is always a (local) minimum of the free
energy. Therefore an (at least metastable) metallic solution is
present even at zero field. The existence of a (local) minimum is a
necessary condition for the occurrence of an hysteretic behavior at
the transition. Of course the region of the hysteresis also depends on
the height of the free energy barrier between the maxima, of domain walls and
of other non-equilibrium properties.  Nevertheless the region in T and
H where two minima exist provides an (excess) estimate for the
hysteresis region experimetally observed in half-doped
Pr$_{0.5}$Sr$_{0.5}$MnO$_3$,
Nd$_{0.5}$Sr$_{0.5}$MnO$_3$,
(Nd$_{1-y}$Sm$_y$)$_{0.5}$Sr$_{0.5}$MnO$_3$ and
Pr$_{0.5}$Ca$_{0.5}$MnO$_3$ \cite{tokura,tokura2}.

\begin{figure}
  \centerline{\resizebox{7cm}{!}{\includegraphics{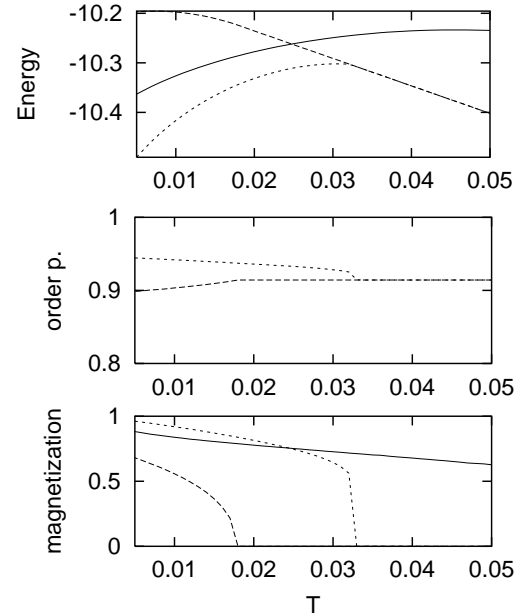}}}
\caption{Energy, order parameter (density) and magnetization vs
  temperature at zero magnetic field. Full line: FM, dashed line: PI,
  dotted line: CE. The parameters correspond to the case PrCaMnO of
  figure 3. In the CE phase, the magnetization is staggered.}
\end{figure}

\begin{figure}
  \centerline{\resizebox{7cm}{!}{\includegraphics{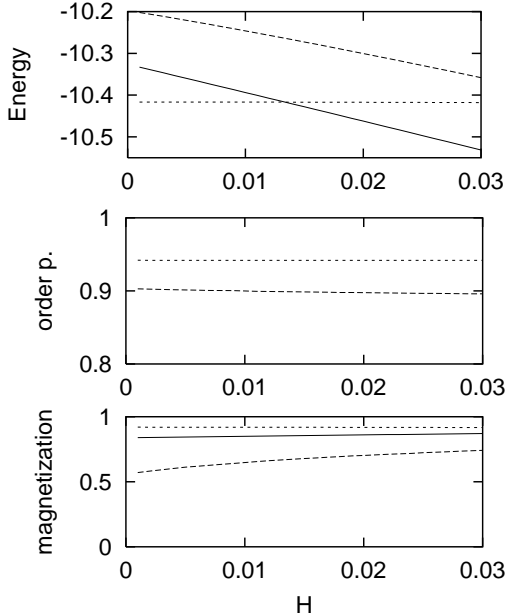}}}
\caption{Energy, order parameter (density) and magnetization vs magnetic
  field at $T=0.01$. Full line: FM, dashed line: PI,
  dotted line: CE. The parameters correspond to the case PrCaMnO of
  figure 3. In the CE phase, the magnetization is staggered.}
\end{figure}

\subsection{Discussion}
The previous subsection illustrated the main results of the present
work:

i) The CE phase does arise in the present one-orbital model and is
crucially related to competition between the JT and breathing
distortions involved in the charge-ordered state on $Mn^{3+}$ and
$Mn^{4+}$ sites respectively.  The shear lattice deformation results
from this competition and couples to the magnetic degree of freedom,
merely through orbital ordering.

ii) Exchange-correlation corrections are essential to stabilize a
metallic ferromagnetic phase, due to substantial screening of both
Coulomb and electron-lattice interactions.

iii) The kinetic energy is a most effective parameter in determining
the relative stability of the various phases upon varying the
temperature and the magnetic field.

As far as point i) is concerned, in the present work we show that the
CE phase not only arises from electronic mechanisms based on the
presence of (at least) two orbitals per Mn site.  The existence of a
CE phase in a model {\em without} orbital degrees of freedom is quite
remarkable. It is indeed repeatedly stated in the literature
\cite{solovyev,jackeli,okamoto} that the CE phase is stabilized by the
kinetic energy gain arising from the orbital ordering forming
ferromagnetic chains. Our results are not in contrast with this
viewpoint, but underline that the above purely electronic mechanism is
not the only possible one, and that the coupling with lattice degrees
of freedom can be of primary importance. In this regard our results
are related to previous Hartree-Fock \cite{mizokawa} and quantum
Monte-Carlo \cite{yunoki} calculations, where the JT deformations were
claimed to be relevant for the occurrence of a CE phase.  Our
low-temperature phase diagrams are qualitatively similar to the one
reported in Fig. 2(c) in Ref.\cite{yunoki} once the distinction
between order and disorder in the orbital degrees of freedom is
discarded \cite{notascalaAF}.  Our contribution in this framework is
to show that the lattice shear deformation is a relevant ingredient in
its own even in the absence of cooperative mechanisms due to
electronic or JT-induced orbital ordering.

Regarding point iii) we notice that a systematic analysis of the role
of the hopping is relevant for the general understanding of the
manganites. In the real materials of the general form
R$_{1-x}$A$_x$MnO$_3$ (where R and A are trivalent rare earth and
divalent alkaline earth ions respectively) the bandwidth can be varied
by changing the radius of the perovskite A site (where the R and A
ions are located).  Depending on the averaged ionic radius the bond
angle of Mn-O-Mn deviates from 180$^0$ in the orthorombic lattice. The
smaller the radius of the A site is, the larger is this angle, which
reduces the Mn-O overlap and the effective Mn-Mn hopping. A systematic
experimental analysis of this effect is reported in Ref.
\cite{tokura2}.  The results summarized in Fig. 3 allow for a unified
qualitative description of different half-doped materials.  In
particular the most insulating behavior in Fig. 3(a) is consistent
with the generic features of Pr$_{0.5}$Ca$_{0.5}$MnO$_3$. On the other
hand, the center panel of Fig. 3 shows the same qualitative behavior of
La$_{0.5}$Ca$_{0.5}$MnO$_3$ or (Nd$_{1-y}$Sm$_y$)$_{0.5}$Sr$_{0.5}$MnO$_3$.
Finally the most metallic system in the right panel is a good qualitative
description of La$_{0.5}$Sr$_{0.5}$MnO$_3$.  Nevertheless, it has been
pointed out \cite{note} that a rapid change in lattice constant $K$,
rather than necessarily small changes of $t$, could be the clue for
the very different behaviours of the systems (A,A$^\prime$)$_{0.5}$MnO$_3$. In
our case this would correspond to an abrupt change along a vertical
line in Figs. 2, and would enhance the first-order character of the
transitions.

A semi-quantitative agreement can even be reached. In fact, the pure
double-exchange ferromagnetic critical temperature ($J/t = 0$) is,
from our 2D mean-field calculation, $T_c\simeq 0.085t$.  A 3D estimate
enhances this value by a factor $3/2$ owing to the number of nearest
neighbours, yielding $T_{c}^{DE} \simeq 0.13 t$. For an average value
$t = 0.3 eV$ one gets a transition temperature $\simeq 450K$. It is
reduced by the presence of the antiferromagnetic coupling, for
instance in panel c of Figure 3, in zero field $T_{c} \simeq 0.05t$
thus $\simeq 270K$ in 3D, thus supporting De Gennes' simple mean-field
picture.  Then one obtains in the center panel the value $T_{c}^{CE} \simeq
180K$, and in the left panel $T_{c}^{CE} \simeq 170K$. These values are
reasonable, as compared with experimental ones, in particular one
notices that $T_{c}^{CE}$ is strongly reduced compared to
$T_{c}^{DE}$.  This is due to the competition between the two order
parameters (ferro and antiferro).  Another way to understand it is to
notice that in the CE phase the effective dimensionality is reduced
by chain formation, together with charge localization this reduces the
effective stength of double-exchange.  On the other hand the effective
antiferromagnetic exchange is close to that of stoechiometric
CaMnO$_{3}$ with a $T_{c}$ of $120K$.

We have also systematically investigated the role of the magnetic
field in stabilizing the uniform FM phase.  The typical energy
differences involved in the first-order transitions are so small that
accessible magnetic fields are sufficient to drive the transition from
the insulating to the FM phases.  Specifically, by taking a typical
value of $t \simeq 0.3 eV$, one can see that $H/t = 0.015$ (where $H =
g \mu_{b} S H$) roughly corresponds to ten Teslas.  This value agrees
well with the typical values experimentally used to investigate the
(T,H) dependence of the low-temperature CE insulating phase and the
intermediate-temperature uniform FM phase \cite{tokura,tokura2}.

\section{Conclusion}

Let us compare our approach with other models which have been proposed
to describe the half-doping compounds.  In our treatment, the main
ingredient which is responsible for the CE-type magnetic ordering is
the appearance of a shear deformation, with the consequent
modification of the magnetic coupling along certain directions.  In
ref. \cite{yunoki}, two different orbitals are retained for the $e_g$
electrons, but the shear deformation is absent. In their approach, the
CE order arises because the orbitals prefer to have large overlaps
along certain directions, thus favoring the kinetic energy along
zig-zag chains where the spins are aligned ferromagnetically.  In ref.
\cite{jackeli}, the electron-lattice interaction is absent, and it is
again the anisotropic $e_g$ transfer amplitude of the two-orbital
model which drives the CE state. In both cases, however, the AF
coupling $J\sim 0.1 t$ necessary to achieve the CE state is one order
of magnitude higher than what is estimated from experiments, signaling
that there must be some additional effect contributing to the CE
ordering. Finally, we reiterate that self-consistent screening is
necessary to explain that phases with marked charge order come in very
close competition with metallic phases. We believe that this is a
crucial feature of doped manganites, that further models addressing
coexistence and texturing of those phases at small scales must take
into account.

%
%
%
%

\begin{thebibliography}{}
\bibitem{notascalaAF} Despite the
  qualitative resemblance of our phase diagrams with the one reported
  in Ref.\cite{yunoki}, we also emphasize that our approach considers
  a strong local e-e repulsion. Possibly this major difference is
  responsible for the much lower values of the AF coupling $J_1$ in
  our results in comparison with the rather large values of $J_{AF}$
  in Ref.\cite{yunoki}.
\end{thebibliography}
%

\end{document}